\documentclass[prl,aps,twocolumn,superscriptaddress]{revtex4-1}
\usepackage{latexsym}
\usepackage{graphicx}
\usepackage{verbatim}
\usepackage{multirow}
\usepackage{amsmath}
\newcommand{\beq}{\begin{eqnarray}}
\newcommand{\eeq}{\end{eqnarray}}
\usepackage{mathrsfs}
\usepackage{float}
\usepackage[usenames, dvipsnames]{color}
\usepackage{mathtools}
\usepackage{slashed}
\usepackage{physics}	
\usepackage{graphicx}   
\usepackage{epstopdf}
\usepackage{subfigure}  
\usepackage{hyperref}   
\usepackage{bbold}
\usepackage{wasysym}
\usepackage{tikz}
\usetikzlibrary{decorations.pathmorphing}
\usetikzlibrary{shapes.misc}
\tikzset{cross/.style={cross out, draw=black, minimum size=8*(#1-\pgflinewidth), inner sep=0pt, outer sep=0pt},
cross/.default={1pt}}

\begin{document}

\title{Detecting isotropic density and nematic fluctuations using ultrafast coherent phonon spectroscopy}
\author{Chandan Setty}
\thanks{Corresponding author: csetty@illinois.edu}
\affiliation{Department of Physics, University of Illinois at Urbana-Champaign, Urbana, Illinois, USA}
\author{Kridsanaphong Limtragool}
\affiliation{Department of Physics, University of Illinois at Urbana-Champaign, Urbana, Illinois, USA}
\author{Byron Freelon}
\affiliation{University of Louisville, Department of Physics and Astronomy, Louisville, KY 40208 }
\author{Philip W. Phillips}
\affiliation{Department of Physics, University of Illinois at Urbana-Champaign, Urbana, Illinois, USA}

\begin{abstract}
We propose a theoretical framework for the detection of order parameter fluctuations in three dimensions using ultrafast coherent phonon spectroscopy with long-range interactions. We focus our attention on long wavelength charge density fluctuations (plasmons), and charged nematic fluctuations where the direction of the propagation vector is fixed perpendicular to the plane of anisotropy.  By treating phonons and light classically and decoupling interactions to integrate out the fermionic degrees of freedom, we arrive at an effective theory of order parameter fluctuations about the spatially uniform saddle-point solution. We find that, due to the $(k_x^2-k_y^2) (B_{1g})$ symmetry of the form factor appearing in the vertex, nematic fluctuations couple to light only at fourth order, unlike isotropic density fluctuations which couple at second order. Hence, to lowest order, the interaction between electrons and the electromagnetic field contributes a driving force for plasmon oscillations while it provides a frequency shift for nematic fluctuations. From the resulting coupled harmonic oscillator equations of motion, we argue that ultrafast coherent phonon spectroscopy could be a useful tool to extract and analyze various electronic properties of interest such as the frequency of the collective mode and the coupling between electrons and phonons.  Specific experiments are proposed on the normal state of high $T_c$ systems to observe the frequency shift predicted here resulting directly from orbital ordering (nematic) fluctuations. Our paper presents a new mechanism for generating coherent phonons from long range interactions (``Coherent Long-range Interaction Induced Phonons or CLIIP") that does not require the existence of multiple bands to act as intermediary states for quasiparticles. 
\end{abstract}

\maketitle
\section{Introduction} 
Order parameter fluctuations have thermodynamic and transport signatures ~\cite{Larkin2005}, and can also be detectable by local probes such as  Muon Spin Resonance ($\mu$SR)~\cite{Yaounac1997, Reotier2011}, Nuclear Magnetic Resonance (NMR) and Nuclear Quadrupole Resonance (NQR)~\cite{Moriya1963, Moriya2003}.   
Useful as they may be, all of these measurements are only indirect probes of fluctuations since they are insensitive to both spatial and temporal information. So far, spin and charge order fluctuations have been best probed in the frequency domain through neutron scattering~\cite{Tranquada2007, Dai2015} and inelastic X-ray scattering~\cite{Schulke2007} or momentum resolved electron energy loss spectroscopy~\cite{Mills2013} respectively, as they contain both spatial and dynamical properties. More recently, Raman scattering in the frequency domain has found increased utility due to its ability to probe orbital and charge nematic fluctuations~\cite{Gallais2013, Gallais2016, Zeyher2013} for certain polarization geometries.

While the aforementioned frequency domain measurements have already provided a wealth of information on order parameter fluctuations in a variety of materials, 
the corresponding time-domain measurements are only now being explored with considerable success~\cite{Orenstein2012}. Time domain measurements have the added advantage of observing fluctuating modes directly, provided the temporal resolution is at least equal to the inverse fluctuation scale of the boson. They also enable a direct extraction of the frequency and lifetime of the fluctuating mode, and as we will see below, the coupling to other modes such as phonons. Evidence of charge density wave fluctuations~\cite{Gedik2013, Liang2013, Mihailovic1999, Shen2008, Lupke2004,  VanLoosdrecht2007, Cavalleri2013} as well as indirect indications of spin fluctuations~\cite{Giannetti2015, Venkatesan2000, Walmsley2000} have been reported by several groups. While amplitude modes of the charge density wave seem to be long-lived, the phase modes for a finite momentum vector are overdamped~\cite{Gedik2013} and as a result are barely observable. Hence, the presence of long-lived fluctuating modes is an essential prerequisite for the observation of oscillations in ultrafast time domain spectroscopy. This statement appears facile at first sight$-$the life-time of the fluctuating mode is independent of whether the measurement is done in time or frequency domain, and decay rates measured in one domain would be expected to carry over to the other. This extension, however, is a non-trivial problem since time domain spectroscopy is inherently out of equilibrium. Therefore, well-known equilibrium results such as the decay rate-self energy relationship and Matthiesen's rule do not generally continue to hold in a non-equilibrium setting~\cite{Freericks2017, Freericks2017-Review} except in the weak-pump limit.

The use of coherent phonons in ultrafast spectroscopy to elucidate the electronic ground states in solids has been well documented~\cite{Shah2013, Merlin1996, Merlin1997, Sabbah2007, Kurz1992, Kurz2000}. Typically, in a pump-probe measurement, the pump pulse excites multiple fluctuating modes as the excitation energies lie in a similar energy window. Eventually these modes become coupled to one another, and hence one must disentangle the electronic information from the data~\cite{Leitenstorfer2012}. This is especially true for the high temperature superconductors since the typical energies of a coherent optical phonon is roughly 5 THz which significantly overlaps with the excitation energies of nematic and spin fluctuations. In these materials, coherent phonon oscillations have been generated~\cite{Nakamura2011} and studied both experimentally~\cite{Marsi2009,Sood2012,Shen2017} and theoretically~\cite{Setty2015} for different ground states. However, in spite of the immensely significant role played by fluctuations in the phenomenology of such materials, few theoretical studies have shed light on the role of fluctuating modes on the oscillations, and no general theoretical framework has been laid.
 
In this work, we provide such a framework by the theoretical principles for probing electronic fluctuating modes through ultrafast coherent phonon spectroscopy via the interaction between electronic and lattice modes. Our focus will be on two long wavelength collective modes in the presence of long-range correlations: plasmons and nematic fluctuations where the direction of the propagation vector is fixed perpendicular to the plane of anisotropy. We treat phonons and the electromagnetic field classically, Hubbard-Stratonovichize the interactions, and integrate out the fermionic degrees of freedom to obtain an effective theory of order parameter fluctuations about the uniform saddle-point solution in the disordered phase. Using perturbation theory, we derive the equations of motion for phonons and the collective mode at zero momentum, resulting in a set of coupled differential equations. We find that, due to the $(k_x^2-k_y^2) (B_{1g})$ symmetry of the form factor appearing in the vertex, nematic fluctuations couple to light only at fourth order (two photon, Raman process), unlike isotropic density fluctuations which couple at second order (one-photon, non-Raman process). Hence, to lowest order, the interaction between electrons and the electromagnetic field contributes a driving force for plasmon oscillations while it provides a frequency shift for nematic fluctuations. Finally, we solve the coupled equations of motion and show how one can extract various electronic properties of interest such as the frequency of the collective mode and the coupling between electrons and phonons. Our paper presents a new mechanism for generating coherent phonons from long range interactions (``Coherent Long-range Interaction Induced Phonons or CLIIP") that does not require the existence of multiple bands to act as intermediary states for quasiparticles.  \par
\begin{figure}
\centering
\includegraphics[width=3.5in,height=1.8in]{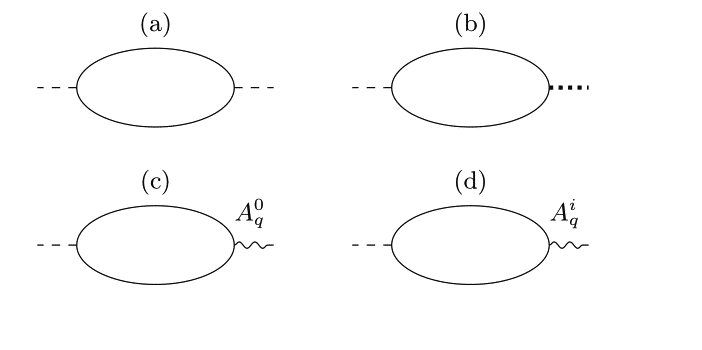}\hfill%
\caption{Feynman diagrams contributing to the effective action for the fluctuating order parameter. The thin solid lines denote free electron Green functions. The dashed, dotted and wavy lines denote the plasmon fluctuating order, phonons and the gauge field respectively. The scalar $A_q^0$ and vector potential $A_q^i$ couple with the electrons with their respective vertices. We have ignored diagrams coupling phonons and the gauge field. } \label{Feynman}
\end{figure}
\section{Plasmons} We begin by recalling the dynamics of zero momentum density fluctuations in three dimensions. The action for electrons with long range Coulomb interactions is given by
\beq \nonumber
S_e\left[\bar{\psi}, \psi \right] &=& \int_0^{\beta} d\tau \bigg[ \int d^3 \textbf{r}~\bar{\psi} \left(\partial_{\tau} + \frac{\hat{p}^2}{2m} - \mu \right)\psi \\
&& + \frac{1}{2} \int d^3\textbf{r}~d^3\textbf{r}'~\bar{\psi}\bar{\psi}'\frac{e^2}{ |\bf r - \bf r' |} \psi'\psi  \bigg] \label{Se}
\eeq
where we have suppressed the indices in the electron Grassmann variables, $\bar{\psi} \equiv \bar{\psi}_{\sigma}(\bf r, \tau)$, and similarly for the primed quantities. Here, $\mu$ is the chemical potential and $\beta$ is the inverse temperature. Using standard many body techniques~\cite{Simons2010}, one can arrive at the effective action for the density fluctuations ($\sigma_q$) about the saddle-point solution $\sigma_q = 0$ given by (in the limit $\frac{v_f |\textbf q|}{\omega}\ll1$, $v_f$ is the Fermi velocity)
\beq
S_e\left[\sigma\right] = \sum_{q} \frac{1}{2} \sigma_q \left( \omega^2 - \omega_p^2 - \frac{3 k_f^2 \omega_p^2}{m^2 \omega^2} |\bf q|^2\right)\sigma_{-q},
\label{sigma}
\end{eqnarray}
where $q$ denotes a collective variable, $q\equiv (iq_n, \bf q)$, containing the Matsubara frequency $iq_n$ and momentum $\bf q$, $\omega_p = \sqrt{\frac{4 \pi n e^2}{m}}$ is the plasmon frequency, $n$ is  the electron density and $k_f$ is the Fermi momentum. We have also used the same notation $\sigma_q$ for the fluctuating field. Note that an expansion in $\frac{v_f |\textbf q|}{\omega}\ll1$ is possible due to the low energy plasmon gap. For a spatially uniform fluctuating order, the equations of motion for $S_e[\sigma]$ yield a simple harmonic oscillator with frequency $\omega_p$. In the presence of an external gauge field and phonons, the action becomes modified to include minimal coupling between electrons and the electromagnetic field, and the electron-phonon interaction. The total action is given by~\cite{Klein1975}
\beq \nonumber
S&=& S_e\left[\bar{\psi}, \psi, A\right] + S_{ph}\left[P,Q\right] + S_{e-ph}\left[\bar{\psi}, \psi, Q\right] \\ \nonumber
S_{e} &=& \int d^3 \textbf{r} d\tau \bigg[ \bar{\psi}\left(\partial_{\tau} - i e A^{0}\right)\psi \\ \nonumber
&&+\frac{1}{2m}\left(\nabla + i e \textbf{A}\right)\bar{\psi}\left(\nabla - i e \textbf{A}\right)\psi - \mu \bar{\psi}\psi \\
&&+ \frac{1}{2} \int d^3\textbf{r}~d^3\textbf{r}'~\bar{\psi}\bar{\psi}'\frac{e^2}{ |\bf r - \bf r' |} \psi'\psi  \bigg] \\
S_{e-ph}  &=&  \sum_{i}\int d^3 \textbf r~\bar{\psi}\psi V(\textbf r - \textbf R_i).
\eeq
Here $A^i$ are the components of the vector potential $A$, $V(r-R_i)$ is the potential coupling the lattice to the electrons, and $S_{ph}\left[P, Q\right]$ is the lattice part of the action which, in momentum space, can be treated as a collection of harmonic oscillators of the form $S_{ph}\left[P,Q\right] = \frac{1}{2}\sum_q \left(P_q^2 + \Omega_q^2 Q_q^2\right)$ ($\Omega_q$ is the frequency of mode $Q_q$ with canonical momentum $P_q$). Henceforth, consistent with previous treatments~\cite{Merlin1996, Merlin1997}, we will also ignore any direct interaction between phonons and light as they coupled indirectly through the electrons. Decoupling the quartic interaction and introducing the Hubbard-Stratonovich field $\phi(\textbf r)$, we rewrite the action as \par
\beq 
 S &=& \int_0^{\beta} d\tau d^3\textbf{r} \bigg[ \frac{1}{8\pi} \left(\partial_{\textbf{r}} \phi \right)^2 + \bar{\psi}\left(G_0^{-1} + \hat{C}(\textbf r)\right) \psi \bigg],
 \label{UnIntegratedAction}
 \eeq
 where
 \beq \nonumber
 \hat{C} &\equiv& \bigg[ -i e A^0 + \frac{i e}{2m} (\textbf A\cdot \nabla + \nabla \cdot \textbf A) + \frac{e^2 |\textbf A|^2}{2m}  \\
 && + i e \phi(\textbf r) + \sum_i V(\textbf r - \textbf R_i) \bigg],
\eeq
and the non-interacting Green function is given by $ G_0^{-1} = \partial_{\tau} -\frac{\nabla^2}{2m} - \mu$. We next treat the electromagnetic field and lattice vibrations classically~\cite{Merlin1996, Merlin1997}, integrate out the fermions, and find the saddle-point solution. We will work in the limit where electrons are only weakly coupled to phonons and the electromagnetic field, and hence the saddle-point solution is not drastically affected. To the zeroth order approximation, this means that we can perform a perturbation expansion for the fluctuations about the original saddle-point solution $\phi_q = 0$. The lowest order terms in the fluctuation expansion contribute to the equations of motion for the phonons and the order parameter field trivially because the terms proportional to the gauge fields are independent of the electron density or lattice fluctuations, and the term proportional to the electron-lattice interaction potential only shifts the zero of the oscillations and can thus be ignored. The second-order terms in the expansion which couple the density fluctuation to light are dictated by gauge invariance and, in the limit of $\frac{v_f |\textbf q|}{\omega}\ll1$, is given by (in terms of the rescaled function $\sigma_q = \frac{1}{\sqrt{4\pi}} \left(\frac{|\textbf q|}{\omega}\right)\phi_q$)
\beq
S_{\sigma-A} = \frac{i \omega_p^2}{\sqrt{4\pi} m} \sum_q \left[ A_q^0 \left( \frac{|\textbf q|}{\omega} \right) \sigma_{-q} + A_q^i \sigma_{-q} \right].
\label{sigma-A}
\eeq
To derive the above equation, we chose a longitudinal component in the electromagnetic pulse to highlight the difference between the nematic fluctuation case (to follow) which has an additional $k_x^2 - k_y^2$ form factor (for a pure transverse pulse the $\sigma-A$ coupling would vanish). In the same limits, the coupling between the density fluctuations to phonons yields
\beq
S_{\sigma-ph} = \frac{i e N_0 v_f^2}{3 \sqrt{4\pi}} \sum_q \xi_q \sigma_{-q} \left(\frac{|\textbf q|}{\omega}\right) Q_q,
\label{sigma-ph}
\eeq
 where $N_0$ is the density of states at the Fermi level and $\xi_q$ is the matrix element for electron-phonon interactions in momentum space. The form of $\xi_q$ is important for the discussions to follow. As we are interested in the zero-momentum transfer limit, the coupling between the density field and phonons for a single band goes to zero unless the matrix element diverges as $\xi_q \sim \frac{1}{|\textbf q|}$. Such a form of the matrix element typically occurs for electrons interacting with lattice displacements through long-range Coulomb interactions. In a multi-band case, zero momentum coupling between electrons and the lattice is allowed even for a constant matrix element. This is because there is a always a non-zero probability amplitude for electrons to be scattered to a different band with zero momentum transfer. Finally, to remain consistent with our previous assumptions of confining ourselves to a minimal model, we ignore the coupling between phonons and the electromagnetic field.  This will have no effect on the order at which density fluctuations contribute.  Under these assumptions, one can collect all the terms contributing to the total action  (Eqs~\ref{sigma}, \ref{sigma-A} and \ref{sigma-ph} along with $S_{ph}$), and determine the equations of motion for $\sigma_q$ and $Q_q$ in the limit of zero momentum transfer ($|\textbf q| \rightarrow 0$, we denote the variables in this limit as $Q_0$ and $\sigma_0$). This leads  to a system of coupled differential equations,
 \beq
 \ddot{\sigma}_0(t) + \omega_p^2 \sigma_0(t) &=& i \gamma^2 A_0^z(t) + i \beta^2 Q_0(t) \label{eom-sigma} \\
 \ddot Q_0(t) + \Omega_0^2 Q_0(t) &=& i \beta^2 \sigma_0(t) \label{eom-Q},
 \eeq
 where the double dots denote second derivatives, $\gamma^2 = \frac{\omega_p^2}{\sqrt{4\pi} m}$, $\beta^2$ is the coefficient of the $1/|\textbf q|$ factor in the electron-phonon coupling matrix element times $\frac{ e N_0 v_f^2}{3 \sqrt{4\pi}}$, and $\Omega_0$ is the frequency of the zero momentum optical phonon. We have also taken the vector potential (the zero momentum component of which is denoted by $A_{0}^i \equiv A_{|\textbf q| = 0}^i$) to point along the $z$ direction. In deriving Eqs~\ref{eom-sigma} and \ref{eom-Q}, we have assumed that close to zero momentum, the frequency of the plasmon is approximately a constant at $\omega\simeq \omega_p$. These equations of motions can be solved for $Q_0(t)$ for a Gaussian pulse with a small width $\tau$ (compared to the inverse frequencies of the individual modes) centered around $t=0$, and we obtain for $t>0$
 \beq
 Q_0(t)= \left[ \frac{e^{-\omega_1^2 \tau^2} sin\omega_1 t}{\omega_1(\omega_1^2 - \omega_2^2)} - \frac{e^{-\omega_2^2 \tau^2} sin\omega_2 t}{\omega_2(\omega_1^2 - \omega_2^2)} \right].
 \eeq
We have chosen the initial conditions such that the amplitude and velocity of the individual modes are zero at $t=-\infty$. The individual frequencies of the oscillation are given by
\beq
\omega_1^2 &=& \frac{1}{2}\left(\Omega_0^2 + \omega_p^2 - \omega'^2\right)\\
\omega_2^2 &=& \frac{1}{2}\left(\Omega_0^2 + \omega_p^2 + \omega'^2\right),
\eeq
where we have defined $\omega'^2 \equiv \sqrt{(\Omega_0^2 - \omega_p^2)^2 - 4 \beta^2}$. Thus, the normalized change in reflectivity of the probe pulse follows (approximately in phase) the modulations caused by the phonon oscillations with the new frequencies $\omega_{1,2}$. These two frequencies can be extracted experimentally from the reflectivity oscillations and, with prior knowledge of the optical phonon frequency ($\Omega_0$, which can be measured from a region of the phase diagram where the fluctuations are small), the frequency of the collective mode ($\omega_p$) and the strength of the electron-phonon coupling ($\beta^2$) can be determined.   

For the simpler case of plasmon oscillations we study in this section, the assumption that the frequency is approximately constant ($\omega \simeq \omega_p$) in the plasmon-phonon coupling (close to zero momentum) can be lifted without much difficulty. In Fourier space, the coupled differential equations of motion for $\sigma_0(\omega)$ and $Q_0(\omega)$ are given by (again in the limit $\frac{v_f |\textbf q|}{\omega}\ll1$)
\beq
-\omega^2 \sigma_0(\omega) + \omega_p^2 \sigma_0(\omega) &=& i\gamma^2 A_0^z(\omega) + i\frac{\beta'^2 Q_0(\omega)}{\omega}, \\
-\omega^2 Q_0(\omega) + \Omega_0^2 Q_0(\omega) &=& -i\frac{\bar{\beta}^2 \sigma_0(\omega)}{\omega}.
\eeq
We have introduced a new electron-phonon coupling constant $\bar{\beta}^2$ to be consistent with the units used previously. The effect of retaining the frequency dependence in the electron-phonon coupling is to force the characteristic polynomial determining the pole structure in $Q_0(\omega)$ to be of third order. Hence, an additional frequency appears superposed on the coherent phonon oscillations. These frequencies, denoted $\bar{\omega_i}$, can be evaluated exactly in the limit of small electron-phonon coupling
\beq
\bar{\omega}_1^2&\simeq& \frac{\beta'^4}{\Omega_0^2 \omega_p^2} \\
\bar{\omega}_2^2 &\simeq& \Omega_0^2 + \frac{\beta'^4}{\Omega_0^2(\Omega_0^2 - \omega_p^2)} \\
\bar{\omega}_3^2 &\simeq& \omega_p^2 + \frac{\beta'^4}{\omega_p^2(\omega_p^2 - \Omega_0^2)}. 
\eeq
Once the three frequencies above are extracted from experiment, $\Omega_0, \omega_p$ and $\beta'^2$ can be evaluated without any prior knowledge of the bare optical phonon frequency. \par
 \begin{figure}
\centering
\includegraphics[width=3.0in,height=1.0in]{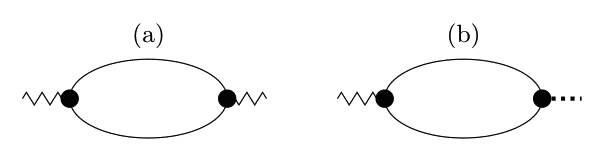}\hfill%
\caption{Feynman diagrams contributing to the effective action at second order for the nematic fluctuations. The thin solid lines denote free electron Green functions. The zig-zag and dotted lines denote the nematic fluctuating order and phonons respectively. The black dots each contribute a $B_{1g}$ form factor $k_x^2 - k_y^2$. Note that we have decomposed the electron phonon interaction in the $B_{1g}$ channel as well, and again ignored diagrams coupling phonons to the gauge field,. } \label{Feynman-second}
\end{figure}
 \section{Nematic fluctuations} As an example of an order parameter fluctuation that couples differently to the electromagnetic field, we consider fluctuations above the nematic ordering transition (in the isotropic phase). Since we are interested in a nematic collective mode in the zero-momentum limit, just as in the case of plasmons, we require that the mode be gapped at $|\textbf q| \rightarrow 0$. This is ensured by the presence of long range interactions in three dimensions that can be decomposed into various irreducible representations of the underlying point group. In the nematic channel for a square lattice, the decomposition will result in a $k_x^2 -k_y^2$ ($B_{1g}$) form factor in the nematic susceptibility~\cite{Gallais2013, Gallais2016}. For simplicity, we will confine ourselves to the case where $\textbf q$ is taken to zero along the axis perpendicular to the anisotropy; although this is not the most general treatment of the problem, our analysis can be easily extended to the case where the momentum transfers are in the plane of the anisotropy. To see that the nematic fluctuations are gapped at zero momentum transfer, we have to evaluate the nematic susceptibility~\cite{Gallais2016}
 \beq
 \pi^n_q = \frac{1}{V} \sum_{\textbf k} f_{\textbf k, \textbf q}^2 \frac{n_f( \epsilon_{\textbf k + \textbf q} ) - n_f( \epsilon_{\textbf k} )}{i\omega_m + \epsilon_{\textbf k + \textbf q} - \epsilon_{\textbf k}},
 \label{nematicsusceptibility}
 \eeq
 where $2 f_{\textbf k, \textbf q} = \left[(k_x^2 - k_y^2) + (k_x + q_x)^2 - (k_y + q_y)^2\right] $, $V$ is the volume and $\epsilon_{\textbf k}$ is the dispersion. Similar to the Lindhard function, $\pi^n_q \sim |\textbf q|^2$ for $|\textbf q| v_f\ll \omega$, and hence,  long range interactions yield a gap in the fluctuation spectrum. Moreover, due to the long-range character of the interactions, even though we work in the continuum limit, the nematic collective mode (like the plasmons) is undamped. This is in contrast with known results~\cite{Fradkin2001, Kivelson2003} where, in the presence of screening, damping effects dominate in the isotropic phase, and hence, collective mode oscillations become effectively indetectable. In addition, recent experiments~\cite{Blumberg2016-arXiv, Blumberg2016-PRB} in the high temperature superconducting pnictides detect a well defined collective mode as the temperature is lowered, indicating that long-range interactions or lattice effects could be important in keeping the modes sharp and detectable by femto-second spectroscopy (an alternate viewpoint has been adopted by the authors of Ref.~\cite{Schmalian2015} where vertex corrections dress the $d$-wave bubble and give rise to a peak that gets sharper when approaching the nematic instability). \par
  \begin{figure}
\centering
\includegraphics[width=3.7in,height=2.8in]{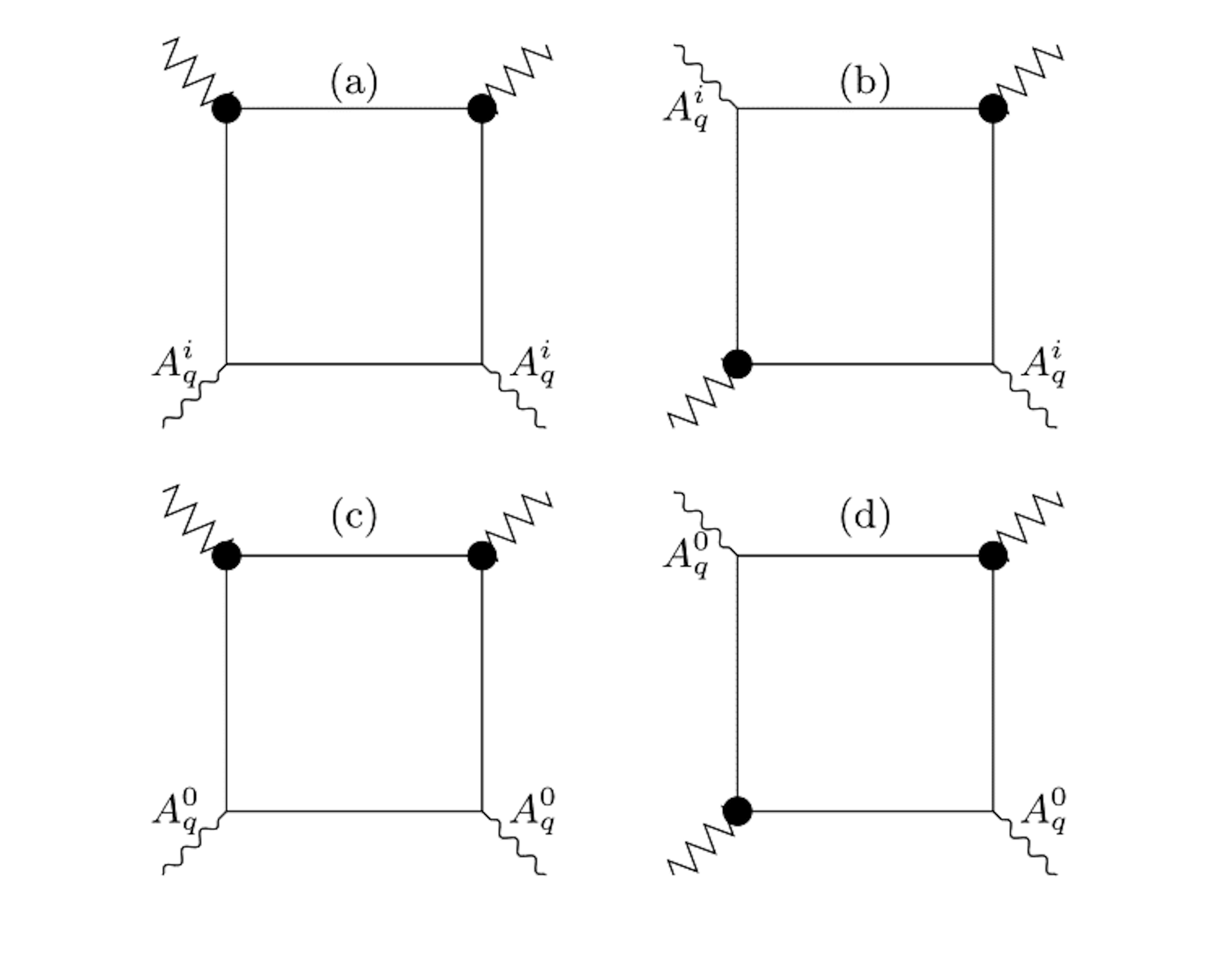}\hfill%
\caption{Feynman diagrams contributing to the effective action at fourth order for the nematic fluctuations. The thin solid lines denote free electron Green functions and the zig-zag lines denote the nematic fluctuating order. The black dots each contribute a $B_{1g}$ form factor $k_x^2 - k_y^2$. The scalar $A_q^0$ and vector potential $A_q^i$ couple with the electrons with their respective vertices. } \label{Feynman-fourth}
\end{figure}
 To write the action for the nematic fluctuations, we replace the interaction term in Eq~\ref{Se} with (see~\cite{Fradkin2001})
 \beq
 S^n_e =  \frac{1}{2} \int d^3(\textbf{r,r'})\bar{\psi}\bar{\psi}'F(\textbf r - \textbf r') \bar{F}(\partial_{x,x'}^2, \partial_{y,y'}^2) \psi'\psi,
 \eeq
 where $F(\textbf r-\textbf r')$ contains the long-range part of the interaction and $\bar{F}(\partial_{(x,x')}^2, \partial_{(y,y')}^2)$ contains the $B_{1g}$ form factors. 
In momentum space, we take $F(\textbf q) \propto 1/q^2 $ with the proportionality constant equal to the  $d$-wave ($l=2$) component of the effective two-body interaction. We set this constant equal to $4 \pi e_n^2$ in analogy with the Coulomb interaction. Note, however, that while the long-range $l=2$ component of the effective potential could have a possible relation to the Coulomb potential,  there is no microscopic origin for the screening of a $d$-wave bubble by the bare Coulomb interaction. We now follow the same procedure outlined in the previous section for plasmons. After performing the Hubbard-Stratonovich transformation and expanding in powers of fluctuations about the uniform saddle point, the second-order term (see Fig~\ref{Feynman-second}(a)) contributing to the effective action for the nematic fluctuation order parameter $\sigma_q^n$ (the superscript $n$ denotes a nematic order parameter fluctuation) is given by (in the limit $\frac{v_f |\textbf q|}{\omega}\ll1$)) 
 \beq
 S_e^n\left[\sigma^n\right] = \sum_{q} \frac{1}{2} \sigma_q^n \left( \omega^2 - \omega_n^2 - \frac{v_f^2\omega_n^2 |\textbf q|^2}{\omega^2} \right)\sigma_{-q}^n,
\label{nematic}
 \eeq
 where $\omega_n$ is the nematic gap given by $\omega_n = 8\pi e_n^2 N_n v_f^2$, $N_n = \frac{m k_f^5}{(2\pi)^2}\frac{16}{105}$, and $k_f$ is the Fermi wave vector. In the same limit, the coupling between phonons and $\sigma_q^n$ is evaluated as (see Fig~\ref{Feynman-second}(b))
 \beq
S_{\sigma-ph}^n = i e_n \sqrt{4\pi} v_f^2 N_n  \sum_q \xi_q \sigma_{-q}^n \left(\frac{|\textbf q|}{\omega}\right) Q_q.
\label{nematic-ph}
\eeq
Note that to obtain Eq~\ref{nematic-ph}, we decomposed the electron-phonon interaction in the $B_{1g}$ channel as well, introducing two $(k_x^2 - k_y^2)$ form factors, and thus keeping the coupling between the nematic fluctuations and phonons non-zero.   \par
The interaction between electrons and light (and thus between $\sigma^n_q$ and the gauge field), however, is constrained by minimal coupling, and cannot be decomposed into $B_{1g}$ lattice form factors. Thus at second order, we can only have one factor of $(k_x^2 - k_y^2)$ whose average value vanishes in the isotropic phase. The lowest non-zero contribution to the $\sigma_q^n - A_q$ coupling obtains from fourth order terms (see Fig~\ref{Feynman-fourth}; the third order triangle diagrams vanish since they are antisymmetric in the internal momenta). There are two classes of fourth-order terms$-$those where the fields $\sigma_q^n$ and $A^{i,0}_q$ alternate on the vertices and those where they appear together (their various permutations yield the same result). These diagrams are shown in Figs~\ref{Feynman-fourth}(a), (c) and Figs~\ref{Feynman-fourth}(b),(d) respectively, and have different contributions to the effective action. For simplicity, we will choose a gauge where $A^0_q$ is zero and, as discussed before, align the vector potential along the z-axis. For small momentum transfers, the diagrams can be cast in the following form
\begin{widetext}
\beq \nonumber \label{nematic-A}
S_{\sigma-A}^n  &=& \frac{e^2 e_n^2}{4 m^2} \sum_{q_{1..3}} \left\{ A_{q_1}^{z}\sigma_{q_2}^n \sigma_{q_3}^n A_{-q_1-q_2-q_3}^z  \omega_{q_2} \bigg[ b~G_1(\omega_{q_k}) - c~G_2(\omega_{q_k}) \bigg] \omega_{q_3} +  \sigma_{q_1}^{n}A_{q_2}^z \sigma_{q_3}^n A_{-q_1-q_2-q_3}^z  \omega_{q_1} \bigg[ b~K(\omega_{q_k})\bigg] \omega_{q_3} \right\}.\\ 
&& 
 \eeq
 \end{widetext}
Here $G_i(\omega_{q_k})$ and $K(\omega_{q_k})$ (defined in the Appendix) are rational functions of the Matsubara frequencies $\omega_{q_k}$ with a pole like structure, whose exact forms are bulky and less enlightening. The first and second terms are obtained from the two classes of diagrams Figs~\ref{Feynman-fourth}(a) and (b) respectively (since we have set $A_q^0 =0$, (c) and (d) do not contribute). The constants $b=\frac{2 k_f^9}{525 \pi^2}$ and $c=\frac{2 k_f^{11}}{2205 \pi^2}$ are obtained from angular integrals of the loop momenta. To simplify Eq~\ref{nematic-A}, we assume spatially uniform fluctuations and gauge fields, and that the fluctuations have approximately a constant frequency $\sim\omega_n$ as we did for plasmons near $\textbf q\rightarrow 0$ (this approximation cannot be made for the frequency of the gauge field since it fails to conserve energy and typically the pump pulse is broad). Under these approximations, the functions $G_i(\omega_{q_k})$ and $K(\omega_{q_k})$ (which are now functions of only the gauge frequencies) acquire poles at $\omega_{q_i} = \pm \omega_n, \pm 2 \omega_n$. Fourier transforming into time domain, simplifying using the Dirac delta functions, and performing standard contour integrals, we rewrite the contribution from the nematic-gauge field coupling term as $S_{\sigma-A}^n = \int dt~\sigma_0^n(t)^2 A_0^z(t) f(i\omega_n t)$. Here the function $f$ is an oscillatory function of $\omega_n t$, and depends on the strength of the external pulse field (note that one factor of $A_0^z$ appears in  $f(i\omega_n t)$ after performing the Fourier integrals). Thus, unlike the case of plasmons, due to the lowest-order coupling between $\sigma_0^n - A_q$ being quadratic in $\sigma_0^n(t)$, the effect of the fourth-order term is to change the frequency of the collective mode.  Collating all the terms in Eqs~\ref{nematic}, \ref{nematic-ph}, \ref{nematic-A} (along with $S_{ph}$), and re-writing them in time domain, we find that the equations of motion for $\sigma_0^n(t)$ and $Q_0(t)$ are determined by
\begin{figure}[h!]
\includegraphics[width=1.7in,height=1.4in]{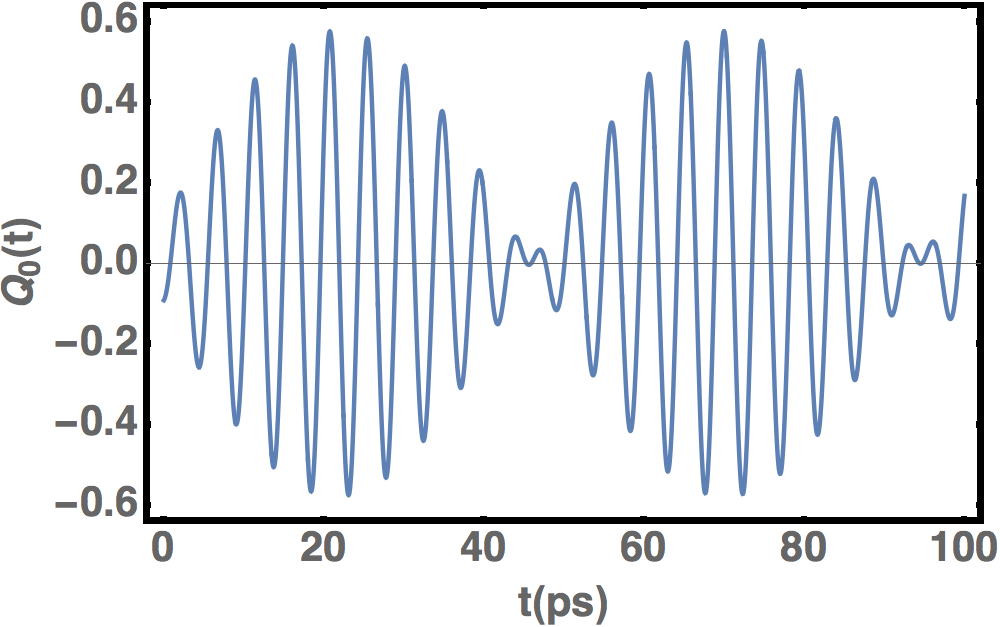}\hfill%
\includegraphics[width=1.7in,height=1.4in]{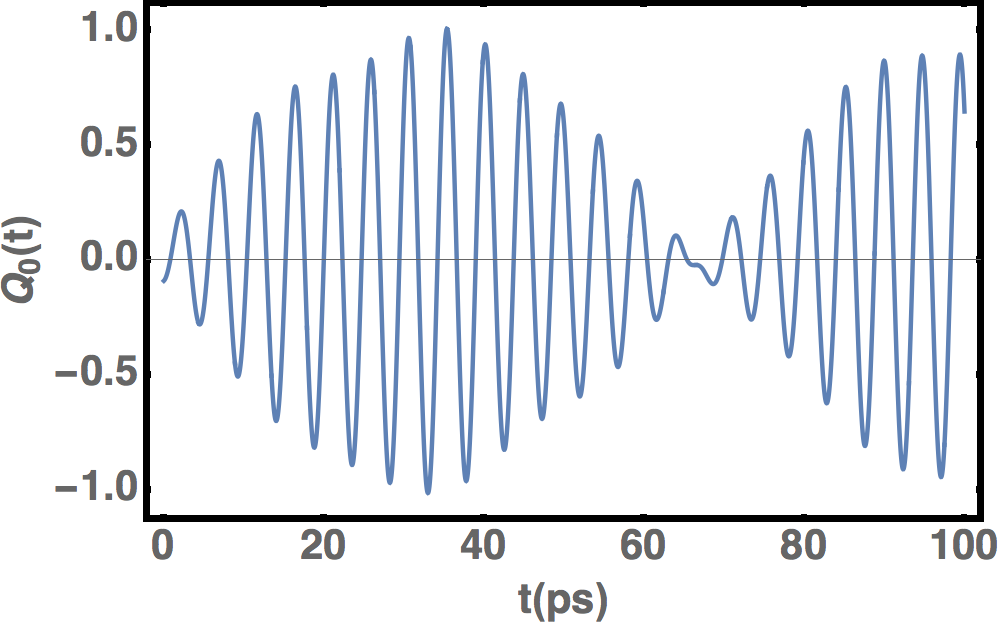}
\caption{Oscillations of the $q=0$ phonon mode as a function of the delay time $t>0$. (Left) For zero coupling ($\gamma'^2=0$) between nematic fluctuations and light and (Right) non-zero coupling $\gamma'^2=0.5$. We have chosen a pulse with a gaussian electric field profile, with a phonon frequency of $\Omega_0=1.26 THz$, nematic mode $\omega_n = 1.41 THz$ and $\beta'^2 = 0.1 THz$. The beat period is increased due to the coupling of nematic fluctuations with light.
} \label{Beats}
\end{figure}
 \beq\label{eom-nematic} 
 \!\!\! \!\!\!\!\!\!\ddot{\sigma}_0^n(t) + \left( \omega_n^2 -\gamma'^2 A_0^z(t) f(i\omega_n t)  \right) \sigma_0^n(t) &=&  i \beta'^2 Q_0(t) \\ 
 \ddot Q_0(t) + \Omega_0^2 Q_0(t) &=& i \beta'^2 \sigma_0^n(t).
 \label{eom-Q-nematic}
 \eeq
Here $\beta'^2$ is given by the product of $e_n \sqrt{4\pi} v_f^2 N_n $ and the coefficient of $1/|\textbf q|$ in the electron-phonon matrix element and $\gamma'^2$ is the effective coupling between $\sigma_0^n(t) - A_0^z$ that depends on the strength of the external field, and is treated as a parameter that can be obtained experimentally. The Eqns~\ref{eom-nematic}, \ref{eom-Q-nematic} can be solved for non-zero initial amplitudes and zero initial velocities for $t<0$ (pulse is incident at $t=0$). When the frequency of the nematic mode is similar to that of the coherent phonons ($\omega_n\sim \Omega_0$, which is typically the case experimentally), one obtains a beating pattern in the phonon oscillations (and hence the change in reflectivity) as shown in Fig~\ref{Beats}. With knowledge of the frequency of the pure coherent phonon oscillations $\Omega_0$, the electronic collective mode frequency ($\omega_n$) and the coupling constant to phonons  ($\beta'^2$) can be extracted in the weak field limit. The effect of $\gamma'^2$, then, is to change the period of the beating pattering by varying the effective frequency of the nematic collective mode. This is depicted in Fig~\ref{Beats} for a simple co-sinusoidal form of $f(i \omega_n t)$, with a pulse having a Gaussian electric field profile. Hence, the coupling between light and nematic collective modes ($\gamma'^2$) can be obtained by changing the intensity of light and observing the shift in the beating pattern. In Fig~\ref{Beats}, the effect of increasing the effective coupling between light and the nematic fluctuations from $\gamma'^2 =0$ (Fig~\ref{Beats}, left panel) to $\gamma'^2 =0.5$ (Fig~\ref{Beats}, right panel) is to increase the beat period by about 20ps.  \par
\section{Discussions} 
At this juncture, it is useful to draw comparisons of our work to existing literature~\cite{Shah2013, Merlin1996, Merlin1997, Sabbah2007, Kurz1992, Kurz2000} on possible mechanisms governing generation of coherent phonons in solids. To this end, we first wish to recall the definitions used by Merlin and collaborators~\cite{Merlin1996, Merlin1997} on two dominant mechanisms that have been proposed in literature -- Impulsive Stimulated Raman Scattering (ISRS) and Displacive Excitation of Coherent Phonons (DECP). In the ISRS mechanism, the force driving the coherent phonons is impulsive, i.e., the width of the excitation pulse $\tau$ is much smaller than the typical inverse frequency of the modes excited. In this case, the coherent phonon displacement field $Q(t)$ follows a simple sinusoidal behavior in time (with no phase shifts). Importantly, the ISRS mechanism involves a three-step Raman scattering process where two photons excite quasiparticles from the ground state band to two consecutive intermediate bands before de-exciting them back to the ground state. The process of de-excitation produces a coherent phonon of the relevant frequency. On the other hand, in a displacive process, the width of the pulse can be comparable to or greater than the inverse frequency of the phonons. In this case, the phonon displacement field $Q(t)$ may have a non-zero phase shift in its oscillation. The analysis of Merlin and co-workers (along with others that followed, see~\cite{Sabbah2007} and references therein) uses a generalized formalism that includes both displacive and impulsive contributions. The general solution for their equations of motion for the phonon displacement field takes the form $Q(t) = exp(-\Gamma t) sin(\Omega_0 t + \phi)$ ($\Gamma $ is a decay factor due to dissipation and $\phi$ is the phase shift) indicating that the dynamics of the modes are displacive when $\phi = \pm \pi/2$ and impulsive when $\phi=0$. In general, their calculations show reasonable fits to experimental data for $\phi$ between 0 and $\pm \pi/2$, i.e., neither displacive or impulsive. With these definitions, we are now in a position to compare and contrast our work with that of Merlin and coworkers and also place our results in the broader context of these definitions existing in current literature. \\ \newline
1) Our model involves only one band, unlike the work of Merlin and collaborators which includes multiple bands. Hence, the three-step scattering process for zero momentum transfer does not hold in our calculations. This makes our work unique as there are no intermediate band states. At this point, a natural question can arise -- in the limit of zero momentum transfer and a single band, the electronic response is expected to be zero and hence there should not be a driving force at all. However, we must note that our main focus is on collective modes (both plasmons and nematic modes) and the presence of long-range interactions is a crucial ingredient that gives a non-zero net response to the driving force. The work of Merlin and coworkers does not include any long range interactions, hence they require multiple bands to yield a non-zero driving force in the equations of motion.   \\ \newline
2) In order to simplify the frequency integrals appearing in their calculations, Merlin and coworkers assume that the incident pulse is very sharply peaked around a central frequency $\omega_0$. This helps with the approximation of frequency integrals to values of the integrand at $\omega_0$. However, this also means that the pulse is very broad in real time. In our calculations, we have chosen the opposite limit by assuming that the pulse is sharply peaked in time at $t=0$ (broad in frequency) and performed the frequency integrals exactly (hence we do not get the additional dependence on central frequency $\omega_0$). In our actual numerics, we used a very small number (much smaller than the inverse phonon frequency) as the temporal width of the pulse.  This gives a non-negligible weight to a very broad range of frequencies,  enough to excite collective modes with smaller energies of THz range. \\ \newline
3) From the two above points, we can infer that our mechanism for coherent phonon generation is neither DECP or ISRS. It is not DECP (displacive) since the temporal width of the pulse is much smaller than the inverse collective mode frequency. It is also not an ISRS mechanism since there are no multiple bands acting as intermediary states for photons. Indeed, our mechanism is a novel way to generate coherent phonon oscillations -- even with a single band -- in the presence of long range interactions (``Coherent Long-range Interaction Induced Phonons or CLIIP"). Since long range interactions form a crucial component in high temperature superconductors through the existence of collective modes, our results could prove useful toward studying properties of collective modes in the Cuprates and Iron superconductors.    \\ \newline
4) Finally, coherent phonon oscillations in the work of Merlin and coworkers are driven purely by the electron-phonon interaction terms in the Hamiltonian. In our work, the driving force is determined by two important ingredients -- electron-phonon interactions just as in the work of Merlin, but also by electron-electron interaction terms yielding the composite collective modes (either plasmons or nematic collective modes induced by long range interactions). This feature couples our equations of motion for the phonons and plasmons/nematic modes and invariably yields a richer behavior.  

\section{Experiments in F\lowercase{e} and C\lowercase{u} High $T_c$ superconductors} 
In typical pump probe measurements, the central frequency of the pulse is about 1.5 eV, while the typical frequencies of interest corresponding to nematic fluctuations is about 50 meV. Hence, for our mechanism to be implemented, one requires pulse widths short enough so that one can access the THz frequency range. This corresponds to pulse widths $< 2$ fs. While obtaining such pulse-widths can be non-trivial with current technology, some state of the art experiments have already produced narrower pulse widths (down to 100 attoseconds), and have successfully applied it in condensed matter systems~\cite{Heinzmann2007}.  \par
The presence of the 1-Fe $B_{2g}$ phonon in the 11 iron superconductor FeS~\cite{Hackl2018} makes it an ideal playground to test our predictions on nematic collective modes. Moreover, diagonal nematics have also been suggested~\cite{Schmalian2018} and observed in the 122 compound BaF$e_2$A$s_2$~\cite{Fisher2016} and the density-wave superconductor HgB$a_2$Cu$O_4$~\cite{Matsuda2018}. In order to utilize the results presented here, several ultrafast techniques could be employed \textit{e.g.,} time-resolved (\textit{tr}) x-ray diffraction (XRD), ultrafast electron diffraction (UED) or \textit{tr}-optical spectroscopy.  To be specific, we suggest that an ultrafast XRD experiment applied to the above systems on a substrate might be an excellent test case.  If a femto-second optical pump signal, is used to excite coherent optical phonons, a subsequent femto-second X-ray pulse could serve to probe coherent phonon spectral behavior through Bragg peak oscillations.  Hence, it should be possible to extract the electron-phonon coupling strength and the nematic collective mode frequency.  In this experiment, the nematic fluctuations should be observed as a shift in the optical phonon oscillations.  We note that the results of this experimental proposal can be compared to the recent pump-probe measurements~\cite{Shen2017} which employed both \textit{tr} XRD and \textit{tr}-ARPES to determine the electron-phonon coupling strength in the other Fe based systems.   \par

\section{Summary}
In summary, we described a minimal model that can be used to detect electronic order parameter fluctuations using ultrafast coherent phonon spectroscopy in three dimensions. We focused on two particular order parameter fluctuations in the zero momentum limit$-$collective (isotropic) modes in the charge density (plasmons), and charge nematic collective modes when the wave vector is taken to zero along a direction perpendicular to the plane of anisotropy (on a square lattice). After performing a perturbation expansion in the fluctuations while treating the electromagnetic field and phonons classically, we derived effective actions for the coupling of the respective collective modes with phonons and the electromagnetic field. Unlike plasmons which couple to light at quadratic order, we found that, due to the $B_{1g}$ form factor in the nematic susceptibility, nematic modes couple to light only at quartic order. Hence, to lowest order, the  coupling between electrons and the electromagnetic field contributes a driving force for plasmons but a frequency shift for nematic fluctuations. We finally determined the equations of motion for the individual collective modes and phonons, and demonstrated how one can extract useful electronic information such as the frequency of the collective mode and electron-phonon/light couplings. Our work provides a first basic theoretical framework for the detection of collective modes using ultrafast coherent phonon spectroscopy, and bears a special relevance to recent time-resolved experiments that have become increasingly popular probes of exotic quantum matter.   \par
 \textit{Acknowledgments:} We acknowledge support from Center for Emergent
Superconductivity, a DOE Energy Frontier Research Center, Grant No. DE-AC0298CH1088.  We also thank the NSF DMR-1461952 for partial funding of this project.
\bibliographystyle{apsrev4-1}
\bibliography{Ultrafast.bib}
\section{Appendix}
In this Appendix, we furnish important details leading up to Eqns~\ref{eom-sigma}, \ref{eom-Q} and Eqns~\ref{eom-nematic}, \ref{eom-Q-nematic}. We begin by integrating out the fermionic degrees of freedom in Eq~\ref{UnIntegratedAction}. The ensuing logarithmic term can be expanded in powers of the perturbation operator $\hat{C}$. As noted in the main text, a strong electromagnetic field will alter the homogeneous saddle point solution from $\phi_q =0$. Since our focus is on fluctuations about the disordered homogenous solution, however, we assume that the electromagnetic field is sufficiently weak so that the saddle point solutions are not altered too much. In this case, the lowest order correction in $\hat C$ gives 
\beq
S^{(1)} = 2 \sum_{k,k'} G_0(k,k')\langle k' \mid \hat C \mid k\rangle
\label{A:FirstOrderTerm}
\eeq
Note that for non-interacting electrons, the Green function is diagonal in momentum. Since we are only interested in the dynamics of the phonon and plasmon fields, all the matrix elements involving the gauge field in Eq~\ref{A:FirstOrderTerm} are unimportant as they do not couple to either $\sigma_q$ and $Q_q$ at first order. Moreover, to first order, electron-phonon coupling term simply shifts the zero of the oscillations trivially, and hence can be ignored. The second order contribution to the action can be written as
\beq \nonumber
S^{(2)} &=& -\sum_{k k'} G_0(k)\langle k\mid \hat C \mid k' \rangle G_0(k') \langle k' \mid \hat C \mid k \rangle \\ \nonumber
\langle k\mid \hat C \mid k' \rangle &=&-i e A_q^0 - \frac{e}{2m}\left( 2 k_i +q_i\right) A_q^i + i e \phi_q \\
&&+ \frac{e^2}{2m} \sum_{p}A^i_{k+q-p}A^i_{p-k} + \xi_{k, k+q} Q_q,
 \eeq
 where we have defined $q = k-k'$ and $i$ is a spatial direction. \par
 \textit{Coupling $\phi_q$ to $A^0_q, A^{(i)}_q$}: We begin with the diagram in Fig~\ref{Feynman} (c) whose contribution is given as
 \beq \nonumber
 S_{\sigma-A^0} &=& (i e)^2 \sum_{k q} G_0(k+q)G_0(k)\left( A_q^0 \phi_{-q} + A_{-q}^0 \phi_q\right) \\ 
 &=&  \frac{(ie)^2}{2}\sum_q A_q^0 \phi_{-q} \left( \pi_q + \pi_{-q}\right)
 \eeq
where, in the limit $|\textbf{q}| \ll k_f$, the free electron polarization function $\pi_q$ is given by 
\beq
\pi_q = -N(0) \left[1 - \frac{i q_n}{2 v_f |\bf q|}Log\left(\frac{iq_n + v_f |\bf q|}{iq_n - v_f |\bf q|}\right)\right].
\eeq
Here $N(0)$ is the density of states at the Fermi level and $v_f$ is the Fermi velocity. Using the inversion properties of $\pi_q$, we can simplify this contribution as
\beq
 S_{\sigma-A^0}= (i e)^2 \sum_q A_q^0~ \pi_q ~\phi_{-q}.
\eeq
We can similarly write the contribution from Fig~\ref{Feynman} (d) as
\begin{widetext}
\beq \nonumber
 S_{\sigma-A^i} &=& \frac{i e^2}{2 m}\sum_{k q} G_0(k+q) G_0(k)\left(2 k_i + q_i\right) \times \left(A_q^i \phi_{-q} + A_{-q}^i \phi_q\right)  =\frac{ie^2}{m}\sum_{kq} A_q^i \phi_{-q} \left( 2 k^i + q^i\right)  \times \left(\frac{n_f( \epsilon_{\textbf k + \textbf q} ) - n_f( \epsilon_{\textbf k} )}{iq_n + \epsilon_{\textbf k + \textbf q} - \epsilon_{\textbf k}}\right),
 \eeq
 \end{widetext}
 which, after performing the momentum integrals, yields (again for and $|\textbf q|\ll k_f$)
 \beq \nonumber
 S_{\sigma-A^i} = \frac{-i e^2 k_f^2}{2 \pi^2} \sum_q A_q^i \phi_{-q}\pi_q'.
  \eeq
Here, we have defined the dimensionless quantity (for $i$ along the $z-$axis perpendicular to the sample plane; we choose a longitudinal component in the electromagnetic pulse to highlight the difference between the nematic case which has an additional $k_x^2 - k_y^2$ form factor)
\beq
\pi_q' = \frac{2iq_n}{v_f q} \left[1 - \frac{i q_n}{2 v_f |\bf q|}Log\left(\frac{iq_n + v_f |\bf q|}{iq_n - v_f |\bf q|}\right)\right].
\eeq
We can now expand $\pi_q$ and $\pi_q'$ in powers of $v_f q/ \omega \ll1$ after performing analytic continuation $iq_n \rightarrow \omega$. Such an expansion is possible since our primary focus is on three dimensions in the presence of long range interactions where the plasmon spectrum is gapped. This gives us
\beq \nonumber
 S_{\sigma-A^0} +  S_{\sigma-A^i} & =& \frac{-e^2 k_f^3}{m \pi^2}\sum_q A_q^0 \frac{|\textbf q|^2}{3 \omega^2} \phi_{-q} \\
 &+& \frac{ie^2 k_f^3}{m\pi^2} \sum_q A_q^i \frac{|\textbf q|}{3 \omega} \phi_q.
 \eeq
 Rescaling the fields $\phi_q = \sqrt{4\pi} \sigma_q \frac{\omega}{|\textbf q|}$, and defining the plasma frequency $\omega_p^2 = \frac{4 p_f^3 e^2}{ 3\pi}$ we obtain Eq~\ref{sigma-A} of the main text for $S_{\sigma -A} \equiv  S_{\sigma-A^0} +  S_{\sigma-A^i}$
 \beq
S_{\sigma-A} = \frac{i \omega_p^2}{\sqrt{4\pi} m} \sum_q \left[ A_q^0 \left( \frac{|\textbf q|}{\omega} \right) \sigma_{-q} + A_q^i \sigma_q \right].
\eeq
 \textit{Coupling $\phi_q$ to $Q_q$}: We now consider the perturbative (second-order) coupling between phonons and the density fields. This term follows similar to the coupling between $\phi_q$ and $A_q^0$ as there is no dependence of the matrix elements on the internal energy-momentum $k$. Hence we can write the contribution of this term to the action as 
 \beq
 S_{\sigma-ph} = i e \sum_q \xi_q \phi_{-q} Q_q \pi_q,
 \eeq
 where $\xi_q$ is the electron-phonon matrix element that depends purely on the energy-momentum transfer. As we will show below, in the zero-momentum transfer limit, the coupling between the order parameter and phonons for a single band goes to zero unless the matrix element diverges as $\xi_q \sim \frac{1}{|\textbf q|}$. Such a form of the matrix element typically occurs for electrons interacting with lattice displacements through long-range Coulomb interactions. Substituting for $\pi_q$ and taking the limit $v_f |\textbf q |\ll \omega $, we obtain
 \beq
  S_{\sigma-ph} = ieN(0) \sum_q \xi_q \phi_{-q} \left(\frac{v_f^2 |\textbf q|^2}{3 (i q_n)^2}\right)Q_q.
 \eeq
 Performing analytic continuation and rescaling $\phi_q = \sqrt{4\pi} \sigma_q \frac{\omega}{|\textbf q|}$ like before, we obtain (Eq~\ref{sigma-ph})
 \beq
S_{\sigma-ph} = \frac{i e N_0 v_f^2}{3 \sqrt{4\pi}} \sum_q \xi_q \sigma_{-q} \left(\frac{|\textbf q|}{\omega}\right) Q_q.
\eeq
Having derived the various plasmon-gauge field and plasmon-phonon coupling terms, we can combine them with the well known contribution of  $S_e[\sigma]$ (diagram Fig~\ref{Feynman} (a), Eq~\ref{sigma}) to obtain the full action. \\
To determine the equations of motion, we write the total effective action $S[\sigma, Q]$ in the zero momentum limit (in this limit we can approximate the frequency denominators by the plasmon frequency $\omega_p$. It is easy to lift this assumption and derive a more general set of solutions as described in the main text) to obtain
\begin{widetext}
\beq
S[\sigma, Q] = \int d\omega \Bigg[ \sigma_0(\omega)(\omega^2 - \omega_p^2) \sigma_0(-\omega) + Q_0(\omega)(\omega^2 - \Omega_0^2)Q_0(-\omega) + i\gamma^2 A_0^z(\omega)\sigma_0(-\omega) + i \beta^2 \sigma_0(-\omega) Q_0(\omega) \Bigg],
\eeq
\end{widetext}
where we have defined $\gamma^2 = \frac{\omega_p^2}{\sqrt{4 \pi} m}$, and $\beta^2$ is the electron-phonon coupling constant (the pre-factor of $1/|\textbf{q}|$ in $\xi_q$ described in the main text). We can now Fourier transform the action into time domain and write the resulting Lagrangian as
\begin{widetext}
\beq
\mathscr{L}(t) = \left( \dot{\sigma}_0(t)^2 - \omega_p^2 \sigma_0(t)^2\right) + \left( \dot{Q}_0(t)^2 - \Omega_0^2 Q_0(t)^2\right) + i \gamma^2 A_0^z(t) \sigma_0(t) + i \beta^2 \sigma_0(t) Q_0(t),
\eeq
\end{widetext}
where the dots on top of the variables denote time derivatives. From the above Lagrangian, it is straightforward to derive the equations of motion to obtain  Eqs~\ref{eom-sigma} and \ref{eom-Q}. \par
\textit{Nematic fluctuations:} To proceed with this calculation, we must first show that in three dimensions and in the presence of long-range interactions, the fluctuation spectrum is gapped at zero momentum. To this end, we must evaluate the integrals in Eq~\ref{nematicsusceptibility}, and show that the lowest order term in the limit of $v_f |\textbf q |\ll \omega $ goes as $|\bf q|^2$. Fixing the direction of momentum transfer perpendicular to the anisotropy (the integral is not the same along other directions), we can write the nematic susceptibility for small momenta in angular coordinates as (the azimuthal angle picks up a factor of $\pi$)
\beq\nonumber
\pi^n(\textbf{q}, i\omega_n) &\simeq& -\frac{1}{8\pi^2} \int k_f^7 |\textbf{k}'|^7 d|\textbf{k}'| sin^5\theta cos\theta d\theta \\
&&\times \left[ \frac{2 \epsilon_f |\textbf{q}'|\delta(\epsilon_{\textbf{k}'} - \epsilon_f)}{i\omega_n + 2 \epsilon_f |\textbf{k}'| |\textbf{q}'| cos\theta} \right].
\eeq
Here $\epsilon_f$ is the Fermi energy, $\theta$ is the vertical angle, and $|\textbf{k}'|$ is the momentum variable defined with respect to the Fermi momentum. The angular integration can be performed to yield after analytic continuation
\beq
\pi^n(\textbf{q}, i\omega_n) \simeq \frac{16 m k_f^5}{105(2\pi)^2} \left(g^2+ \frac{g^4}{3} + ..\right),
\eeq
where $g\equiv \frac{v_f |\textbf{q}|}{\omega}$. Hence, in the presence of long range interactions, the fluctuation spectrum is gapped with a gap value $\omega_n^2= 8\pi e_n^2 N_n v_f^2$ where $N_n = \frac{m k_f^5}{(2\pi)^2}\frac{16}{105}$ (see Ref~\cite{Chubukov2018} for other interesting cases).\par
\textit{$\phi_q^n - \phi_q^n$ and $\phi_q^n - Q_q$ couplings}: We are now in a position to evaluate the diagrams appearing in Figs~\ref{Feynman-second} and \ref{Feynman-fourth}. The calculation of diagrams in Figs~\ref{Feynman-second}(a) and (b) proceeds analogous to the case of plasmons, but with the insersion form factors $f_{\textbf k, \textbf q}^2$ into the susceptibility. With this modification, the electronic contribution from Fig~\ref{Feynman-second}(a) to the effective action becomes (after replacing the expression for $\pi^n(\textbf q,i\omega_n)$)
\beq \nonumber
S_{e}^n = \sum_q \phi_q^n \left[\frac{|\textbf{q}|^2}{8\pi} - e_n^2 N_n \left( \frac{v_f^2 |\textbf{q}|^2}{\omega^2} + \frac{v_f^4 |\textbf{q}|^4}{3 \omega^4} \right) \right]\phi_{-q}^n.
\eeq
Substituting for the nematic collective mode energy $\omega_n^2$ and rescaling the fields $\sigma_q^n \equiv \frac{\phi_q^n}{2}\sqrt{\frac{|\textbf{q}|^2}{4\pi \omega^2}}$, we obtain (Eq~\ref{nematic} of the main text)
\beq
S_{e}^n = \frac{1}{2} \sum_q \sigma_q^n \left(\omega^2 - \omega_n^2 -\frac{\omega_n^2 v_f^2|\textbf{q}|^2 }{\omega^2} \right) \sigma_{-q}^n.
\eeq
Similarly, the contribution from the diagram in Fig~\ref{Feynman-second}(b) takes the form 
\beq
S_{\sigma-ph}^n = ie_n \sum_q \xi_q ~\phi_{-q}~\pi_q^n~Q_q,
\eeq
which, to lowest order in $g$ becomes
\beq
S_{\sigma-ph}^n \simeq ie_n N_n \sum_q \xi_q ~\phi_{-q}~g^2~Q_q. 
\eeq
Rescaling $\phi_{-q}^n = \sqrt{4\pi} \frac{\omega}{|\textbf q|}\sigma_{-q}^n $, we obtain Eq~\ref{nematic-ph} of the main text. \par
\textit{Coupling $\phi_q^n$ to $A^0_q, A^i_q$}: Before we begin, we note that only diagrams that contain an even number of fluctuation vertices contribute. This happens because, in the isotropic phase, an odd number of form factors $f_{\textbf k, \textbf q}$ have a zero momentum average (these terms would contribute in the symmetry broken phase). Furthermore, the third order triangle diagrams vanish since they are antisymmetric in the internal momenta. We demonstrate the main steps in the calculation for coupling $\phi_q^n$ to $A^i$ (see Figs~\ref{Feynman-fourth}(a) and (b)). We work in the gauge where $A^0_q$ is zero, hence we are left with the two types of diagrams in Figs~\ref{Feynman-fourth}(a) and (b), which we denote as type (a) and type (b) respectively. The contribution of these diagrams to the effective action is given by
\begin{widetext}
\beq \nonumber
 S_{\sigma-A}^{n(a)}&=& \frac{e^2 e_n^2}{4 m^2}\sum_{k q_1 q_2 q_3} f_{k}^2 (2 \textbf k +  \textbf q_1 )_i (2 \textbf k +  \textbf q_1 +  \textbf q_2 + \textbf q_3)_i \phi_{q_2} \phi_{q_3} A_{q_1}^i A_{-q_1 -q_2-q_3}^i \\ 
 && \times G_0(k) G_0(k+q_1) G_0(k+q_1 + q_2)G_0(k+q_1 + q_2 +q_3) \\ \nonumber
 S_{\sigma-A}^{n(b)}&=& \frac{e^2 e_n^2}{4 m^2} \sum_{k q_1 q_2 q_3} f_{k}^2 (2 \textbf k + 2 \textbf q_1 + \textbf q_2 )_i (2 \textbf k +  \textbf q_1 +  \textbf q_2 + \textbf q_3)_i \phi_{q_1} A_{q_2}^i \phi_{q_3} A_{-q_1 -q_2-q_3}^i \\ 
 && \times G_0(k) G_0(k+q_1) G_0(k+q_1 + q_2)G_0(k+q_1 + q_2 +q_3)
\eeq
The Fermionic Matsubara sum over $i k_n$ can be performed over the product of Green functions, which after a variable shift, gives (at $T=0$)
\beq \nonumber
\sum_{i k_n}G_0(k) G_0(k+q_1) G_0(k+q_1 + q_2)G_0(k+q_1 + q_2 +q_3) &=&\\  \nonumber
&& \!\!\!\!\!\!\!\!\!\!\!\!\!\!\!\!\!\!\!\!\!\!\!\!\!\!\!\!\!\!\!\!\!\!\!\!\!\!\!\!\!\!\!\!\!\!\!\!\!\!\!\!\!\!\!\!\!\!\!\!\!\!\!\!\!\!\!\!\!\!\!\!\!\!\!\!\!\!\!\!\!\!\!\!\!\!\!\!\!\!\!\!\!\!\!\!\!\!\!\frac{1}{(\epsilon_{\bf k} - \epsilon_{\bf k+ q_1} + i\omega_{q_1} )(\epsilon_{\bf k} - \epsilon_{\bf k+ q_1 + q_2} + i\omega_{q_1} + i \omega_{q_2} )(\epsilon_{\bf k} - \epsilon_{\bf k+ q_1 + q_2 +q_3} + i\omega_{q_1} + i\omega_{q_2} + i\omega_{q_3})} \\ \nonumber
 &&\!\!\!\!\!\!\!\!\!\!\!\!\!\!\!\!\!\!\!\!\!\!\!\!\!\!\!\!\!\!\!\!\!\!\!\!\!\!\!\!\!\!\!\!\!\!\!\!\!\!\!\!\!\!\!\!\!\!\!\!\!\!\!\!\!\!\!\!\!\!\!\!\!\!\!\!\!\!\!\!\!\!\!\!\!\!\!\!\!\!\!\!\!\!\!\!\!\!\!+ \frac{1}{(\epsilon_{\bf k} - \epsilon_{\bf k+ q_1} - i\omega_{q_1}  )(\epsilon_{\bf k} - \epsilon_{\bf k - q_2} + i \omega_{q_2}  )(\epsilon_{\bf k} - \epsilon_{\bf k- q_2 -q_3} + i \omega_{q_2}  + i\omega_{q_3} )} \\ \nonumber
 &&\!\!\!\!\!\!\!\!\!\!\!\!\!\!\!\!\!\!\!\!\!\!\!\!\!\!\!\!\!\!\!\!\!\!\!\!\!\!\!\!\!\!\!\!\!\!\!\!\!\!\!\!\!\!\!\!\!\!\!\!\!\!\!\!\!\!\!\!\!\!\!\!\!\!\!\!\!\!\!\!\!\!\!\!\!\!\!\!\!\!\!\!\!\!\!\!\!\!\!+  \frac{1}{(\epsilon_{\bf k} - \epsilon_{\bf k+ q_1 +q_2} - i\omega_{q_1}  - i\omega_{q_2}  )(\epsilon_{\bf k} - \epsilon_{\bf k - q_2} - i \omega_{q_2}  )(\epsilon_{\bf k} - \epsilon_{\bf k -q_3} + i\omega_{q_3} )} \\ \nonumber
 && \!\!\!\!\!\!\!\!\!\!\!\!\!\!\!\!\!\!\!\!\!\!\!\!\!\!\!\!\!\!\!\!\!\!\!\!\!\!\!\!\!\!\!\!\!\!\!\!\!\!\!\!\!\!\!\!\!\!\!\!\!\!\!\!\!\!\!\!\!\!\!\!\!\!\!\!\!\!\!\!\!\!\!\!\!\!\!\!\!\!\!\!\!\!\!\!\!\!\!+ \frac{1}{(\epsilon_{\bf k} - \epsilon_{\bf k+ q_3} - i\omega_{q_3}  )(\epsilon_{\bf k} - \epsilon_{\bf k+ q_2 + q_3} - i\omega_{q_2}  - i \omega_{q_3}  )(\epsilon_{\bf k} - \epsilon_{\bf k+ q_1 + q_2 +q_3} - i\omega_{q_1}  - i \omega_{q_2}  - i\omega_{q_3} )}.
\eeq
Here, $\omega_{q_i}$ are the corresponding Matsubara frequencies for the energy-momenta $q_i$, and $\epsilon_{\textbf k}$ is the bare dispersion. We can now make the substitution $\epsilon_{\textbf k} - \epsilon_{\textbf k + \textbf {q}_i} = -(2 |\textbf k| |\textbf{q}_i| cos \theta_{\textbf k, \textbf{q}_i} + |\textbf q|_i^2 ) $, $\epsilon_{\textbf k} - \epsilon_{\textbf k + \textbf {q}_i +\textbf {q}_j} = -(2 |\textbf k| |\textbf{q}_i + \textbf {q}_j|  cos \theta_{\textbf k, \textbf{q}_i + \textbf{q}_j} + (\textbf q_i + \textbf {q}_j)^2 ) $, where $\theta_{\textbf k, \textbf{q}_i}$ is the angle between the momenta $\textbf k$ and $\textbf{q}_i$. For small momentum transfers, we can expand the energy denominators up to second order in $\textbf q_i$, and arrange the terms in powers of $\bf q_i \bf q_j$. These momenta can be used to rescale the $\phi_q$ fields in terms of the $\sigma_q$ fields as was done in the previous paragraphs; the remaining terms vanish in the limit of zero momentum transfer. The resulting $\bf k$ integrals are then straightforward to perform and yield (defining $S_{\sigma-A}^n \equiv S_{\sigma-A}^{n (a)}+S_{\sigma-A}^{n (b)}$ and choosing the polarization along the z-direction)
\beq \nonumber
S_{\sigma-A}^n  &=& \frac{e^2 e_n^2}{4 m^2}\sum_{q_{1..3}} \left\{ A_{q_1}^{z}\sigma_{q_2}^n \sigma_{q_3}^n A_{-q_1-q_2-q_3}^z  \omega_{q_2} \bigg[ b~G_1(\omega_{q_k}) - c~G_2(\omega_{q_k}) \bigg] \omega_{q_3} +  \sigma_{q_1}^{n}A_{q_2}^z \sigma_{q_3}^n A_{-q_1-q_2-q_3}^z  \omega_{q_1} \bigg[ b~K(\omega_{q_k})\bigg] \omega_{q_3} \right\}.\\ 
&& \label{nematic-A-App}
 \eeq
 where we have defined
\beq \nonumber
G_1(\omega_{q_i})&=&\frac{2 \omega_{q_1}}{\omega_{q_2}^2 \left( \omega_{q_1} + \omega_{q_2}\right)^2 \omega_{q_3} } + \frac{3}{\omega_{q_2} \left( \omega_{q_1} + \omega_{q_2}\right)^2 \omega_{q_3}} - \frac{3 \omega_{q_2} + \omega_{q_3}}{\omega_{q_1}\omega_{q_2}^2 \left(\omega_{q_2} + \omega_{q_3}\right)^2} \\
&&+ \frac{6 \left(\omega_{q_1} + \omega_{q_2}\right) + 2 \omega_{q_3}}{\omega_{q_1}\left( \omega_{q_1} + \omega_{q_2}\right)^2\left( \omega_{q_1} +\omega_{q_2} +\omega_{q_3}\right)^2} \\ 
G_2(\omega_{q_i})&=&\frac{-4}{\omega_{q_2}^2 \left( \omega_{q_1} + \omega_{q_2} \right) \omega_{q_3}^2} \\ 
K(\omega_{q_i})&=& \frac{\omega_{q_1}^4 + 2 \omega_{q_1}^3 \omega_{q_2} + \omega_{q_2}\omega_{q_3} (\omega_{q_2} + \omega_{q_3})^2 + 2 \omega_{q_1}\omega_{q_3}(\omega_{q_2}+ \omega_{q_3}) (4 \omega_{q_2} + \omega_{q_3}) + \omega_{q_1}^2 (\omega_{q_2}^2 + 7 \omega_{q_2}\omega_{q_3} + 7 \omega_{q_3}^2) }{\omega_{q_1}^2 (\omega_{q_1} + \omega_{q_2} )^2 \omega_{q_3}  (\omega_{q_2}  + \omega_{q_3} ) (\omega_{q_1}  + \omega_{q_2}  +\omega_{q_3} )^2}.
\eeq
These are the quantities that appear Eq~\ref{nematic-A} of the main text. For simplicity, we choose a spatially uniform order parameter and gauge field; this simplifies the spatial components of the momentum integrals. Moreover, since we are restricting to small momenta, we can set the frequency of the collective mode to be approximately a constant at the value of the gap i.e.,  $\omega_n$ (note that this cannot be done for the frequency of the electromagnetic field) just as we did for plasmons. Fourier transforming into real time and summing over the residues (poles occur at $\pm \omega_n, \pm 2 \omega_n$), we are left with two integration variables giving (by using the resulting Dirac delta functions)
\beq
S_{\sigma-A}^n = 12 \pi \frac{e^2 e_n^2}{4 m^2} \int dt dt' \omega_n \sigma(t)^2 A^z(t') A^z(t) \left( sin\omega_n\tilde t  +  sin 2\omega_n\tilde t + \omega_n \tilde t~ cos \omega_n \tilde t + \omega_n\tilde t ~cos 2 \omega_n \tilde t  \right)
\eeq
\end{widetext}
where $\tilde t = t -t'$. Writing the oscillating functions in terms of exponentials, one can perform a Fourier transform for a general $A^z(t')$ with respect to the variable $t'$. The remaining terms oscillate with respect to the variable $t$ at frequencies $\omega_n, 2\omega_n$, and define the function $f(i\omega_n t)$ that was discussed in the main text. We can therefore write the net contribution to the action from nematic fluctuation-gauge field coupling as $S_{\sigma-A}^n = \int dt~\sigma_0^n(t)^2 A_0^z(t) f(i\omega_n t)$. Note that, implicit in the definition of $f(i\omega_n t)$ above, is the magnitude of $A^z(t')$ that is present in the integration over $t'$. We have included the parameter $\gamma'$ in the main text to reflect this dependence. Hence, one can change the magnitude of the applied field to control the beat pattern frequency. Collecting all the terms, $S_e^n, S_{\sigma-ph}^n, S_{\sigma-A}^n $ together, one can derive the equations of motion Eq~\ref{eom-nematic} and \ref{eom-Q-nematic} in the main text.
\end{document}